# On the Higgs mechanism and the gauge field theory

**V M Koryukin**
 Mari State University, Lenin square - 1, Yoshkar-Ola, 424001, Russia.

E-mail: kvm@marsu.ru

**Abstract.** As laboratory experiments for the detection of particles with non-zero rest masses forming the dark matter do not give positive results we offer once more to turn the attention upon the neutrinos background of the Universe. If the neutrinos background has the temperature 2 K, then direct observations of particles are impossible ones and only their high density allows hope for the success indirect observations. In consequence of this the field theory is constructed as the maximum plausible reduction of the Feynman formulation of the quantum theory displaying experiment data adequately.

## 1. Introduction

At present it is come to its close the enthusiasm concerned with the detection of vector bosons $W^+$, $W^-$, $Z^\circ$ which are responsible for the weak interaction and are predicted by theoretically. In the first place we note that not a fundamental scalar particle (for example: the Higgs boson, the axion [1]) is detected although there is the set of the composite scalar (pseudoscalar) mesons (hadrons with spin 0). What is more we do not know anything about the graviton (the fundamental tensor boson with spin 2) which must be responsible for the gravitational interaction in the quantum theory by definition. At the same time there are a major number of the observed fundamental fermions with spin 1/2, if we shall relate the color quarks together with leptons to this class. Thereby experimental data hint at the existence of the fundamental Universe property concerned with the spin nature of elementary particles which must be laid in the foundations of physical theory.

Quite possibly that by the construction of theory it is necessary to reverse the concept of vacuum regarding hadrons as holes in the quark sea of particles being in the ground (degenerate) state by the temperature $T_\circ \sim 10^{-13}\,GeV$ (we shall use the system of units $h/(2\pi) = c = 1$, where h is the Planck constant and c is the velocity of light), the estimation of which may be the temperature of the cosmic microwave background detected by Penzias and Wilson in 1964 [2]. As it is known the fact, what the vacuum energy density of electromagnetic field was the negative one, awaken the hope at Casimir to construct the model of the extended elementary particles in the form of spheres in which the Coulomb repulsion is placed in equilibrium with the attraction, connected with zero-point oscillations of vacuum [3]. The Casimir energy, calculated by the computer, proved the positive one and was equal to $E = +0{,}09235/(2r)$ ($r$ is the sphere radius) [4]. Thus it had to seek the other causes explaining a stability of elementary particles what causes us to remember the Dirac hypothesis 1930 about the existence of the electron sea with the negative energy. In our opinion the degenerate fermions pressure of the ground (vacuum) state can be the main cause, not allowing a fermion, having the charge, "is inflated".

The offered hypothesis by us explains both the confinement (within hadrons the confinement of color quarks which's we shall consider as exited states of ground fermions) and the asymptotic freedom (within hadrons at small distances (where the effective temperature is high enough) the conduct of color quarks as freedom particles). In the degenerate state ground fermions of Universe, generating Fermi and Bose liquids, are weakly-interacting particles, but it is not excluded by the interaction with hadrons their exhibition as color fermions – ghosts [5]. We do not exclude also the possibility, that in the form of the Fermi liquid they must be considered as right neutrinos and left antineutrinos with the sufficiently high Fermi energy. It must be exhibited in the absence of these particles by weak interactions of low energies (a mirror asymmetry). Thus for example, it can be interpreted a lepton production upon a charged pion decay as a freezing-out of color degrees of freedom what is expressed in the form of the spontaneous breaking of the $SU(3)$ symmetry characterizing the interaction of color quarks to the $SU(2) \times U(1)$ symmetry characterizing the electroweak interactions of leptons.

Naturally, that the color plasma frequency, defined by the energy of the compound scalar boson having the minimal mass, allows to give the crude estimation of the density $n_{\circ} \sim m_{\pi}^{2} m_{q} \sim 10^{-3} GeV^{3}$ ($m_{\pi} \sim 10^{-1} GeV$ is a mass of pion, $m_{q}$ is a mass of light quark ($m_{q} \sim 10^{-2} \div 10^{-1} GeV$)) ground fermions of Universe.

As stated above in spite of all attempts fundamental scalar particles was unable to detect (supporters of an existence of scalar Higgs bosons await the starting time of the large hadron collider at CERN due at 2008, but it is necessary to note that they have the justification for the negative outcome even if in the subsequent increase of this particles rest masses) and all familiar scalar particles are compound ones (scalar mesons). This problem attracts an attention more and the more so, that the Lagrangian of scalar fields in the standard electro-weak model has obviously the macroscopic nature which is necessary for the actuation of the mechanism of the spontaneous breaking of symmetry. The similar mechanism [6] of the spontaneous breaking of symmetry was considered in the theory of superconductivity for Cooper pairs and it would be logically to use pairs from ground fermions of Universe instead of scalar particles in standard electro-weak model. It can be remove questions on the infinitesimal of the interaction constant of the charged fermions with the scalar field and its dependence from their flavors [7].

In the elementary particles physics the spontaneous breaking of symmetry, realized by compound fields and having the name of the dynamic breaking of symmetry, was considered with 1961 in works of many authors (the literature on this subject can be taken in papers [8,9]). In connection with this it must be noted the work [9], in which in our opinion are made the important commentaries on the role of the quark condensate in order to make hadrons massive ones. Let us note that the mass problem in the gauge theory of arbitrary interactions was the urgent one only for three fundamental particles: $W^{+}$, $W^{-}$, $Z^{\circ}$. The photon and gluons are massless particles. The other bosons are compound ones and the entire problem with masses transfers to fermions for which's it do not exist simply. In consequence of this it was necessary to decide the question on the mass nature only for $W^{+}$, $W^{-}$, $Z^{\circ}$ bosons what we are doing, advancing the hypothesis on the major background of Universe particles which's exist in the ground state and which's manifest itself in the weak interaction [10].

The large density of particles in the Universe interacting only weakly is confirmed also by the considerable value ($\sim 10^{2} GeV$) of rest masses $m_{W}$ and $m_{Z}$ accordingly of $W^{\pm}$-bosons and $Z^{\circ}$-boson generating the weak interaction. Here for a vacuum we have the analog of the first-kind superconductor with the large coherence length (its role can be played the value $1/H_{\circ}$, giving the estimation to the length of a free run of particle in "vacuum" (the Hubble constant $H_\circ \sim 10^{-42} GeV$ )) and with the small London penetration depth of the weak field (the value $1/m_Z$ can play its role) [11]. Using the analog of the known formula [12] for the London penetration depth of the magnetic field ($\lambda_L^2 = m_p c^2 / \left(4\pi n_p g^2\right)$, where $\lambda_L$ is the London penetration depth, $m_p \sim 10^{-10} \div 10^{-9} GeV$ is the mass of the Cooper pair (we consider that $m_p \sim m_\nu$, where $m_\nu$ is the supposed rest mass of the electron neutrino), $g \sim 10^{-2} \div 10^{-1}$ is the charge of the Cooper pair), we can do the crude estimation of the density of the Cooper pairs: $n_p \sim n_\circ \sim 10^{-3} GeV^3$.

Moreover it can obtain this estimation of density $n_\circ$ using the kinetic relation connecting the mean free path of charged particles in a vacuum with the density of Universe ground neutrinos, if we shall take into account only the weak interaction. We shall use the known empirical relation [13] $H_\circ / G_N \approx m_\pi^3$ considering that the Hubble constant (also as before) gives the estimation $1/H_\circ$ the mean free path $l \sim 1/(n_\circ \sigma_\nu)$ of a particle in a vacuum ($\sigma_\nu$ is the scattering cross-section of a neutrino on a charged particle) and taking into account the interpretation given before for the gravitation constant $G_N \sim 10^{-38} GeV^{-2}$ ($G_N \sim \sigma_\nu \propto \alpha G_F^2 T_\nu^2$ [14], $\alpha \sim 10^{-2}$ is the fine structure constant, $G_F \sim 10^{-5} GeV^{-2}$ is the Fermi constant, $T_\nu \sim 10^{-13} GeV$ is the temperature of Universe ground neutrinos). Let us note what in this case it can not assume giant fluxes of high energy neutrinos from astrophysical sources for the explanation of the observable flux of events of cosmic rays of ultrahigh energies within the scope of the *Z-burst* mechanism (the *Z-burst* mechanism is the mechanism of the generation of cosmic rays of ultrahigh energies in a result of an annihilation of a high energy neutrino (under which we shall imply antineutrinos, too) on an neutrino background of the Universe) [15].

The Hubble constant may be connect also with the shielding length of the electromagnetic field in a vacuum because of the photon scattering by virtual charged particles the high density of which's is supported by the presence of Universe ground particles. As is known [16] the last astronomical data caused to introduce such notion as the dark energy endowing some of matter properties (energy, pressure) on the bare vacuum in addition to the available dark matter which was introduced before for the explanation of the non-standard behavior of galaxies [1]. This notion (and instead of it the more careful notion – the quintessence) was called to conserve the standard Friedman model of the expanding Universe, but many its problems remain not solved up to now. Specifically it is the planeness problem (the space curvature close to zero) and the horizon problem (the high degree of isotropism of cosmic microwave background from causal free regions of the Universe) [1]. In the Einstein theory because of the gravitational instability of a matter with a low density this facts are difficult to explain.

In our opinion it necessary to make the more radical decision, namely to support Hubble doubts [17] in regard to the explanation of the red frequency shift of an electromagnetic radiation of distant galaxies at the expense of its recession (Hubble assumed the photon ageing) and to return to the construction of the stationary model of the Universe. As was said before for this it is necessary to revise cardinally the matter quantity, considering that the greater its part exist in the degenerate state, characterized by the low temperature. If it will adopt for its estimation the temperature of the cosmic microwave background in the Universe, then for this we receive sufficient basis. Moreover namely the role of irreversible processes in the Universe become principal one by the construction its stationary model. It allows looking at new on Hoyle and Narlikar hypothesis [18] on the dependence of elementary particle masses on space-time coordinates.

## 2. The maximum plausible functions

It is obvious that a fundamental problem solution of an arbitrary physical system description is run into information incompleteness about a Universe matter. It causes us to make use of a probability interpretation of field functions for a production of a model-independent solution of its problem. Naturally that for this aim the Feynman formulation [19] of the quantum theory is more attractive one. As a result the classical description must satisfy the requirement of the maximum probability.

We note that in the twentieth century it was made largely attempts to break down the determinism which firmly established in the science to the nineteenth century close. Certainly the quantum mechanics inserted the principal contribution in this, the formation of which was initiated by results of experiments in atomic and nuclear physics. But and in the base of bases in which the determinism was founded - in the classical mechanics - it was marked "disadvantages" causing to the loss of illusions [20]. Despite the fact that with illusions were finished it is difficult to give up the determinism idea, as the planning of the physical experiments was based on the accounts relying on the methods which appeared in the science during the determinism domination.

In the first place to these methods it is necessary to attribute the infinitesimal calculus. It is difficult to overestimate the successes in the field. We can indicate only at a field of mathematics - the Lie group theory - which exercised the huge influence on all theoretical physics. Of course, here the useful results can be received owing to the "good" properties of the used spaces (it is used the Hausdorff spaces in spite of the quantum nature of laws acting in the microcosm). What is more the availability of the smooth congruence's which are the solutions of differential equations plays the important role in the Lie group theory. At the same time in the quantum mechanics the existence of the elementary particles paths are negated. In consequence of this instead of the Lie derivatives it is becoming necessary to use the more general operators which can induce the more general algebraic structures in comparison with the Lie groups. Specifically it can be the Lie local loops [21] which's allow to take into account the absence of the determinism in the real physical processes.

Note that the description non-adequacy of the physical systems by the smooth fields in the differentiable manifolds lead to the necessity to give the probabilistic interpretation to the geometrical objects. In consequence of this we shall consider the solutions of the differential equations only as the maximum likelihood functions used for the description of these systems. Of course by this we take account of laws acting in the microcosm and regard them more the fundamental ones than those which are used for the description of macroscopic bodies motions.

We shall rely on the approach suggested by Schrödinger [22] which introduced the set of the unorthogonal to each other wave-functions describing the unspreading wave packet for a quantum oscillator. Later Glauber [23] showed a scope for a description of coherent phenomena in the optics by the Schrödinger introduced states and it was he who called them as coherent ones. This approach received the further development in Perelomov's work's who proposed the definition of the generalized coherent states specifically as the states arising by the action of the representation operator of a some transformation group on any fixed vector in the space of this representation [24]. It is what allows giving the physical interpretation to the gauge transformations by our opinion as the transformations inducing the generalized coherent states, which are characterized by the continuous and possibly hidden parameters [25].

Let us to consider the packet $\{\Psi(\omega)\}$ of functions and let the substitutions

$$\Psi \to \Psi + \delta\Psi = \Psi + \delta T(\Psi) \tag{1}$$

are the most general infinitesimal ones where $\delta T$ are infinitesimal operators of a transition. We draw smooth curves through the common point $\omega \in M_r$ with the assistance of which we define the corresponding set of vector fields $\{\delta\xi(\omega)\}$. Further we define the deviations of fields $\Psi(\omega)$ in the point $\omega \in M_r$ as

$$\delta_\circ\Psi = \delta X(\Psi) = \delta T(\Psi) - \delta\xi(\Psi) \qquad (2)$$

and we shall require that these deviations were minimal ones even if in "the mean". If we state the task – to find the smooth fields $\Psi(\omega)$ in the studied domain $\Omega_r$ of the parameters space $M_r$ then it can turn out to be unrealistic one (possibly $r >> 1$ and possibly $r \rightarrow \infty$). That's precisely therefore the task of the finding of the restrictions $\Psi(x)$ on the manifold $M_n$ ($x \in M_n \subset M_r$, $n \le r$).

Let the square of the semi-norm $|X(\Upsilon)|$ has the form as the following integral

$$A = \int\limits_{\Omega_n} \Lambda\, d_n V = \int\limits_{\Omega_n} \kappa \overline{X}(\Psi)\rho X(\Psi)d_n V \;. \qquad (3)$$

(we shall name A as an action and $\Lambda$ as a Lagrangian also as in the field theory). Here and further $\kappa$ is a constant; $\rho = \rho(x)$ is the density matrix ( $\mathrm{tr}\rho = 1$, $\rho^+ = \rho$, the top index "+" is the symbol of the Hermitian conjugation) and the bar means the generalized Dirac conjugation which must coincide with the standard one in particular case that is to be the superposition of Hermitian conjugation and the spatial inversion of the space-time $M_4$. Solutions $\Psi(x)$ (and even one solution) of equations, which are being produced by the requirement of the minimality of the integral (20) can be used for the construction of the all set of functions $\{\Psi(x)\}$ (generated by the transition operator).

Of course for this purpose we can use the analog of the maximum likelihood method employing for the probability amplitude, but not for the probability as in the mathematical statistics. As is known [19], according to the Feynman's hypothesis the probability amplitude of the system transition from the state $\Psi(x)$ in the state $\Psi'(x')$ equal to the following integral

$$K(\Psi, \Psi') = \int\limits_{\Omega(\Psi, \Psi')} \exp(iA)D\Psi =$$

$$\lim_{N \to \infty} I_N \int d\Psi_1 ... \int d\Psi_k ... \int d\Psi_{N-1} \exp\left( i \sum_{k=1}^{N-1} \Lambda(\Psi(x_k))\Delta V_k \right) \qquad (4)$$

($i^2 = -1$; the constant $I_N$ is chosen so that the limit is existing). Therefore the functions $\Psi(x)$, received from the requirement of the minimality of the action A, are also the maximum likelihood ones only. In this approach the Lagrangian $\Lambda$ plays the more fundamental role than differential equations which are received from it.

Let $\eta_{i_1 ... i_n}(x)$ are the basic $n$-vector components of the affine connection space $M_n$ [26], $\Gamma_{ij}^k(x)$ are the components of the internal connection (here and further Latin indices $i, j, k, ...$ will run the values of integers from 1 to $n$) , $\nabla_i$ is the symbol of the covariant derivative in regard to the connection $\Gamma_{ij}^k(x)$. We shall consider the domain $\Omega_n \subset M_n$ bounded by the hypersurface $\Omega_{n-1}$ covered by one map. Let us to introduce just one more the symbol of the covariant derivative $\overset{\circ}{\nabla}_i$ in regard to the connection $\overset{\circ k}{\Gamma_{ij}}(x)$ given on $\Omega_n$ such what $\overset{\circ}{\nabla}_i \eta_{j_1 \ldots j_n} = 0, \overset{\circ k}{\Gamma_{ij}} = \overset{\circ k}{\Gamma_{ji}}$.

We write the generalization Stokes formula [27]:

$$\oint_{\Omega_{n-1}} W_{i_1 \ldots i_{n-1}} d\tau^{i_1 \ldots i_{n-1}} = \int_{\Omega_n} \partial_{i_n} W_{i_1 \ldots i_{n-1}} d\tau^{i_1 \ldots i_{n-1} i_n}, \tag{5}$$

where $W_{i_1 \ldots i_{n-1}}(x)$ are the tensor field components, $d\tau^{i_1 \ldots i_{n-1}}$ is the form of the hypersurface $\Omega_{n-1}$ volume and $d\tau^{i_1 \ldots i_{n-1} i_n}$ is the form of the domain $\Omega_n$ volume. Let $T_i = T_{ij}^j$, where the components $T_{ij}^k$ of the affine strain tensor [26] are defined in the form $T_{ij}^k = \Gamma_{ij}^k - \overset{\circ k}{\Gamma_{ij}}$ and besides let $W^{i_n} = W_{i_1 \ldots i_{n-1}} \eta^{i_1 \ldots i_{n-1} i_n}$. As [27]

$$d\tau^{i_1 \ldots i_n} = \eta^{i_1 \ldots i_n} d_n V, \qquad d\tau^{i_1 \ldots i_{n-1}} = \eta^{i_1 \ldots i_{n-1} i_n} N_{i_n} \varepsilon(N) d_{n-1} V, \tag{6}$$

where $N_{i_n}(x)$ is the normal to the hypersurface $\Omega_{n-1}$, $\varepsilon(N) = \pm 1$ is the indicator of the vector $N_{i_n}$, $\eta^{i_1 \ldots i_n}$ is the $n$-vector being the relative one to the basic $n$-vector, and

$$\eta^{i_1 \ldots i_{n-1} i_n} \partial_{i_n} W_{i_1 \ldots i_{n-1}} = \eta^{i_1 \ldots i_{n-1} i_n} \overset{\circ}{\nabla}_{i_n} W_{i_1 \ldots i_{n-1}} = \overset{\circ}{\nabla}_{i_n} W^{i_n} = \partial_{i_n} W^{i_n} + \overset{\circ}{\Gamma}_{i_n}^{i_n} W^{i_n},$$

then the formula (5) can be written in the form

$$\oint_{\Omega_{n-1}} W^{i_n} N_{i_n} \varepsilon(N) d_{n-1} V = \int_{\Omega_n} (\nabla_i W^i - T_i W^i) d_n V. \tag{7}$$

We received the generalization of Green formula in a non-homogeneous space, which will be used for the derivation of field equations and conservation laws.

Let $E_{n+N}$ is the vector fiber space with the base $M_n$ and with the projection $\pi_N$, $\Psi(x)$ is the section of the vector bundle $E_{n+N}$, $\nabla_i$ is the symbol of the covariant derivative in regard to the connections $\Gamma_{ij}^k(x)$ and $\Gamma_i(x)$. Let us to consider a variation of an action A in regard to following arbitrary infinitesimal substitutions:

$$x \to x, \quad \Psi \to \Psi + \delta_\circ \Psi, \quad \nabla_i \Psi \to \nabla_i \Psi + \delta_\circ \nabla_i \Psi,$$

where

$$\delta_\circ \nabla_i \Psi = \nabla_i \delta_\circ \Psi - T_i \delta_\circ \Psi, \tag{8}$$

by this the changes $\delta_\circ \Psi(x)$ of functions must be reduced in zero on a bound $\Omega_{n-1}$ of a domain $\Omega_n$. Let a Lagrangian $\Lambda$ depend on covariant derivatives of fields $\Psi(x)$ not over the first order. As a result a variation of an action is written in the form:

$$\delta_\circ A = \int\limits_{\Omega_n} \left( \frac{\partial \Lambda}{\partial \Psi} \delta_\circ \Psi + \frac{\partial \Lambda}{\partial \nabla_i \Psi} \delta_\circ \nabla_i \Psi \right) d_n V =$$

$$\int\limits_{\Omega_n} \left( \frac{\partial \Lambda}{\partial \Psi} - \nabla_i \left( \frac{\partial \Lambda}{\partial \nabla_i \Psi} \right) \right) \delta_\circ \Psi d_n V + \int\limits_{\Omega_n} \left( \nabla_i \left( \frac{\partial \Lambda}{\partial \nabla_i \Psi} \delta_\circ \Psi \right) - T_i \frac{\partial \Lambda}{\partial \nabla_i \Psi} \delta_\circ \Psi \right) d_n V. \tag{9}$$

Using the formula (7) we transform the last integral of the expression (9) in the integral on a surface which will equal to zero in consequence of a conversion in zero changes $\delta_\circ \Psi(x)$ on this surface. Thus for arbitrary changes $\delta_\circ \Psi(x)$ of functions in a domain $\Omega_n$ a variation $\delta_\circ A$ of the action A can be converted in zero only by the following condition:

$$\frac{\partial \Lambda}{\partial \Psi} - \nabla_i \left( \frac{\partial \Lambda}{\partial \nabla_i \Psi} \right) = 0. \tag{10}$$

Solving given equations it can receive the maximum plausible functions $\Psi(x)$ describing the states of physical systems induced by operators of a transition $\delta T(\Psi)$. The total packet of functions can be received only with the formula (1).

### 3. The realizations of Lie local loops

In order to obtain a classical theory the field number must be a minimum one and a sufficient one at the same time for a description of a closed physical system naturally in limits of parameter changes which's are possible ones for tests. Precisely in consequence of the latter it's enough to use the Lie local loop of infinitesimal substitutions of field functions.

The set $Q$, considered together with a binary operation $\beta$, is named as the groupoid and is labeled by $Q(\beta)$. The groupoid $Q(\beta)$ is named as the quasi-group, if the equations: $\beta(a,x) = b$, $\beta(y,a) = b$ are solvable for any $a, b \in Q(\beta)$ always and what is more in a unique fashion [28]. The quasi-group $Q(\beta)$ with the unit is named as the loop, that is in the loop there is an element $e \in Q(\beta)$ such that for any elements $a, b \in Q(\beta)$ are satisfied identities: $\beta(e,a) = a$, $\beta(b,e) = b$. The loop and the group will be labeled by the symbol $G$. If for any three elements from $G$ the correlation

$$\beta(\beta(a,b),c) = \beta(a,\beta(b,c)) \tag{11}$$

(the associativity axiom) is satisfied then $G$ is named as the group. In this case the binary operation is written also in the form: $\beta(a,b) = ab$. The one-to-one mapping of the set $Q$ onto oneself is named as the substitution of set $Q$. The operation B is named the isotopic one to the operation $\beta$, or the isotope of $\beta$, if there is the ternary of substitutions $i$, $j$, $k$ of set $Q$ such that $B(x,y) = k^{-1} \beta(ix, jy)$ for any $x, y \in Q(\beta)$. There is the <u>theorem</u>: *Each quasi-group is isotopic to some loop* [28].

The loop $Q(\beta)$ will be named as the local one if the binary operation $\beta$ is given for some pairs of elements from a neighbourhood of unit only. If in addition to it, $Q$ is the differentiable manifold and $\beta$ is the differentiable operation then $Q(\beta)$ is named as the local Lie loop. Mal'tsev [21] introduced the notion of the local Lie loop in 1955 year. In 1964 year Kikkawa [29] proofed that in an each point neighbourhood of an affine connection space it can defined the multiplication operation relative to which this neighbourhood becomes the local Lie loop. Let $M$ is the topological space and $G$ is the local loop. We shall name the topological space $M$ as the local quasi-homogeneous one, if the local loop $G(x_\circ)$ acts transitively in $U(x_\circ) \subset M$. If the loop $G$ acts transitively on $M$, then we shall name $M$ as the quasi-homogeneous space [30].

Let $E_{n+N}$ is the vector fiber space with the base $M_n$ and the projection $\pi_N$, $\Psi(x)$ is the arbitrary section of fibre bundle $E_{n+N}$, $\partial_i$ is the partial derivative symbol. Let us to consider the infinitesimal substitutions defining the vector space mapping of the neighbour points $x$ and $x + \delta x$ ($x \in U$, $x + \delta x \in U$, $U \subset M_n$) and holding the possible linear dependence between vectors. We write given substitutions as:

$$\Psi'(x + \delta x) = \Psi(x) + \delta \Psi(x) = \Psi(x) + \delta T(x, \Psi). \qquad (12)$$

By this the vector field change in consequence of the transition in the neighbour point has the form: $\Psi(x + \delta x) - \Psi(x) \approx \delta x^i \partial_i \Psi(x)$ and the change of the field $\Psi$ in the point $x + \delta x$ will equal

$$\delta_\circ \Psi(x + \delta x) = \Psi'(x + \delta x) - \Psi(x + \delta x) =$$
$$= \Psi'(x + \delta x) - \Psi(x) - [\Psi(x + \delta x) - \Psi(x)] \approx \delta T(x, \Psi) - \delta x^i \partial_i \Psi(x).$$

Further we shall denote

$$\delta_\circ \Psi(x) = \delta X(\Psi) = \delta T(x, \Psi) - \delta x^i \partial_i \Psi(x). \qquad (13)$$

Let the formula (12) defines the infinitesimal substitution of the Lie local loop $G_r(x)$ moreover the unit $e$ of the Lie local loop, the co-ordinates of which equal to zero, corresponds to the identity substitution. Then the infinitesimal substitutions of the Lie loop in co-ordinates are written as

$$x^i \to x^i + \delta x^i = x^i + \delta \omega^a(x) \xi_a^i(x), \qquad (14)$$

$$\Psi^A(x) \to \Psi^A(x) + \delta \omega^a(x) T_a{}^A(x, \Psi), \qquad (15)$$

where $x^i$ are the co-ordinates of the point $x$, $x^i + \delta x^i$ are the co-ordinates of the point $x + \delta x$, $\Psi^A(x)$ are the components of the vector field $\Psi(x)$ and $\delta \omega^a(x)$ are the components of the infinitesimal vector field $\delta \omega(x)$ being the section of the vector fibre bundle $E_{n+r}$ with the base $M_n$ and with the projection $\pi_r$ (here and further Latin indices $a$, $b$, $c$, $d$, $e$, $f$, $g$, $h$ will run the values of integers from 1 to $r$ and Latin capital indices $A$, $B$, $C$, $D$, $E$ will run the values of integers from 1 to $N$).

As a result the formula (13) is rewritten in the following form:

$$\delta_\circ \Psi = \delta \omega^a X_a(\Psi), \qquad (16)$$

where $X_a(\Psi) = T_a(\Psi) - \xi_a^i \partial_i \Psi$ or in the co-ordinates $X_a^A(\Psi) = T_a^A(\Psi) - \xi_a^i \partial_i \Psi^A$. In the general case a type of geometrical objects can do not conserving with the similar substitutions. Therefore below we shall consider only such substitutions which are conserving a type of geometrical objects.

In first in the formula (16) it ought to become to the covariant derivative. Let

$$\delta_\circ \Psi^A = \delta\omega^a X_a^A(\Psi) = \delta\omega^a (L_a^A(\Psi) - \xi_a^i \nabla_i \Psi^A), \qquad (17)$$

where

$$L_a^A(\Psi) = T_a^A(\Psi) + \xi_a^i \Gamma_{iB}^A \Psi^B, \qquad \nabla_i \Psi^A = \partial_i \Psi^A + \Gamma_{iB}^A \Psi^B,$$

and we demand that $L_a^A(\Psi)$ and $\xi_a^i(x)$ should be the components of intermediate [26] tensor fields. Hence if $\Psi(x)$ are the components of the vector field then $\Psi(x) + \delta_\circ \Psi(x)$ also are the components of the vector field.

We shall name the fields $X_a(\Psi)$ as the generators of the Lie local loop $G_r(x)$, if the multiplication $[X_a X_b]$ satisfies the following two axioms:

$$[X_a X_b] + [X_b X_a] = 0, \quad [[X_a X_b] X_c] + [[X_b X_c] X_a] + [[X_c X_a] X_b] = 0. \qquad (18)$$

If

$$[X_a X_b] = X_a X_b - X_b X_a = C_{ab}^c X_c. \qquad (19)$$

$$L_a^A(\Psi) = L_{aB}^A \Psi^B,$$

then the intermediate tensor fields $L_{aB}^A(x)$ and $\xi_a^i(x)$ must satisfy to the following correlations:

$$L_{aC}^B L_{bB}^A - L_{bC}^B L_{aB}^A + \xi_a^i \nabla_i L_{bC}^A - \xi_b^i \nabla_i L_{aC}^A - \xi_a^i \xi_b^j R_{ijC}^A = -C_{ab}^c L_{cC}^A,$$

$$\xi_a^i \nabla_i \xi_b^k - \xi_b^i \nabla_i \xi_a^k - 2\xi_a^i \xi_b^j S_{ij}^k = -C_{ab}^c \xi_c^k,$$

where $S_{ij}^k(x)$ are the components of the torsion tensor and $R_{ijC}^A(x)$ are the curvature tensor components of the connection $\Gamma_{iC}^A(x)$. The components $C_{ab}^c(x)$, alternating on down indices owing to (19) of the structural tensor, must satisfy in consequence of (18) to the generalized Jacobi identities

$$C_{[ab}^d C_{c]d}^e - \xi_{[a}^i \nabla_{|i|} C_{bc]}^e + \xi_{[a}^i \xi_b^j R_{|ij|c]}^e = 0 \qquad (20)$$

( $R_{ijc}{}^{e}(x)$ are the curvature tensor components of the connection $\Gamma_{ia}^{b}(x)$ ).

Note that if the Lie local loop $G_r(x)$ operates in the space of the affine connection as transitively so and effectively ($n = r$), then choosing the components $\xi_a^k$ of the intermediate tensor field equaled to the Kronecker symbols $\delta_a^k$ it can show that the correlations (20) become in the Ricci identity [26] when $C_{ab}^c = 2S_{ab}^c$.

**4. Noether theorem and gauge fields**

We demand that the action A was the invariant one with respect to the infinitesimal substitutions (14) and (15) of the local Lie loop $G_r(x)$ conserving the type of geometrical objects. Taking account of the formula (6) we write the $n$-dimensional volume element $d_nV$ in the form

$$d_nV = \eta(x)d_{\circ n}V = \eta(x)dx^1 dx^2 \ldots dx^n .$$

By infinitesimal substitutions (14) the volume element $d_{\circ n}V$ is changed as

$$d_{\circ n}V \rightarrow d_{\circ n}V + \frac{\partial\left(\delta\omega^a \xi_a^i\right)}{\partial x^i} d_{\circ n}V .$$

Let us to consider the corresponding variation of the action (3):

$$\delta\text{A} = \int\limits_{\Omega_n} \left( \left( \frac{\partial\Lambda}{\partial\Psi}\delta_{\circ}\Psi + \frac{\partial\Lambda}{\partial(\nabla_i\Psi)}\delta_{\circ}(\nabla_i\Psi) \right)\eta + \Lambda\eta\partial_i\left(\delta\omega^a \xi_a^i\right) + \partial_i(\Lambda\eta)\delta\omega^a \xi_a^i \right) d_{\circ n}V . \qquad (21)$$

Taking account of the condition (8) we write the expression (21) as:

$$\delta\text{A} = \int\limits_{\Omega_n} \left( \left( \frac{\partial\Lambda}{\partial\Psi} - \nabla_i\left( \frac{\partial\Lambda}{\partial(\nabla_i\Psi)} \right) \right)\delta\omega^a X_a(\Psi) - \nabla_i\left( I_a^i\delta\omega^a \right) + T_i I_a^i\delta\omega^a \right) d_nV ,$$

where

$$I_a^i = -\Lambda\xi_a^i - \frac{\partial\Lambda}{\partial(\nabla_i\Psi)}X_a(\Psi) .$$

In that case when arbitrary parameters $\delta\omega^a$ are constants we receive only the following correlations from the requirement of the action (3) invariance in regard to substitutions (14) and (15) (Noether theorem):

$$\left( \frac{\partial\Lambda}{\partial\Psi} - \nabla_i\left( \frac{\partial\Lambda}{\partial(\nabla_i\Psi)} \right) \right)X_a(\Psi) = \nabla_i I_a^i - T_i I_a^i , \qquad (22)$$

в противном случае мы получаем дополнительно жесткие ограничения на лагранжиан: $I_a^i = 0$.

So that the appearance of stringent restrictions can be excluded on a Lagrangian when $\nabla_i \delta \omega^a \neq 0$ we introduce its dependence on additional fields $B(x)$, the changes of which's contain the covariant derivatives of infinitesimal parameters $\nabla_i \delta \omega^a (x)$, in consequence of this they are named as gauge fields. Let

$$BB^+ = \rho \operatorname{tr}\left(BB^+\right) \tag{23}$$

in the Lagrangian (3).

Further fields $\Psi(x)$ we shall name as prime ones. Considering the factorization on the local Lie loop $G_r(x)$, we denote the components of the gauge fields $B(x)$ as: $B_\alpha^c(x)$ (Greek indices take on values which's not concretized on account of foregoing). By this the transformation low of gauge fields $B(x)$ must have the form:

$$B_\alpha^c \to B_\alpha^c + \delta_\circ B_\alpha^c + \delta \omega^b \xi_b^i \partial_i B_\alpha^c \ , \tag{24}$$

where the changes $\delta_\circ B_\alpha^c$ in a point $x \in M_n$ are defined by the equality

$$\delta_\circ B_\alpha^c = \delta \omega^b U_{b\alpha}^c (B) + \nabla_i \delta \omega^b Z_{b\alpha}^{ic}(B).$$

Here and further $U_{b\alpha}^c (B)$ and $Z_{b\alpha}^{ic}(B)$ are arbitrary functions of gauge fields which's ought to be defined.

Let us to study the action

$$A_t = \int_{\Omega_n} \Lambda_t \left(\Psi, \nabla_i \Psi, B, \nabla_k B \right) \eta(B) d_\circ n V, \tag{25}$$

which we shall note by the total one by substitutions (14), (15), (24). We demand that the action is the invariant one in regard to considering substitutions. Taking account of the parameters arbitrariness $\delta \omega^a$, $\nabla_i \delta \omega^a$, $\nabla_{(i} \nabla_{j)} \delta \omega^a$ and also the identity $T_i + \partial_i \ln \eta - \Gamma_{ki}^k = 0$, we receive the following correlations:

$$\nabla_i \left( \frac{\partial \Lambda_t}{\partial \nabla_i \Psi} X_a(\Psi) + \frac{\partial \Lambda_t}{\partial \nabla_i B_\gamma^c} U_{a\gamma}^c(B) + \Lambda_t \xi_a^i \right) - T_i \left( \frac{\partial \Lambda_t}{\partial \nabla_i \Psi} X_a(\Psi) \right.$$

$$+ \frac{\partial \Lambda_t}{\partial \nabla_i B_\gamma^c} U_{a\gamma}^c(B) + \Lambda_t \xi_a^i \right) + \left( \frac{1}{\eta} \frac{\partial (\Lambda_t \eta)}{\partial B_\gamma^c} - \nabla_i \left( \frac{\partial \Lambda_t}{\partial \nabla_i B_\gamma^c} \right) \right) U_{a\gamma}^c(B) \tag{26}$$

$$+ \left( \frac{\partial \Lambda_t}{\partial \Psi} - \nabla_i \left( \frac{\partial \Lambda_t}{\partial \nabla_i \Psi} \right) \right) X_a(\Psi) + \frac{1}{2} R_{ika}{}^b \frac{\partial \Lambda_t}{\partial \nabla_i B_\gamma^c} Z_{b\gamma}^{kc}(B) = 0,$$

$$\Lambda_t\left(\xi_b^k + Z_{b\alpha}^{kc}(B)\frac{\partial \ln \eta}{\partial B_\alpha^c}\right) + \frac{\partial \Lambda_t}{\partial \nabla_k \Psi}X_b(\Psi) + Z_{b\alpha}^{kc}(B)\frac{\partial \Lambda_t}{\partial B_\alpha^c} +$$

$$\frac{\partial \Lambda_t}{\partial \nabla_i B_\alpha^c}\left(\delta_i^k U_{b\alpha}^c(B) + \nabla_i Z_{b\alpha}^{kc}(B) - T_i Z_{b\alpha}^{kc}(B) - S_{ij}^k Z_{b\alpha}^{jc}(B)\right) = 0, \tag{27}$$

$$\frac{\partial \Lambda_t}{\partial \nabla_i B_\alpha^c}Z_{b\alpha}^{kc}(B) + \frac{\partial \Lambda_t}{\partial \nabla_k B_\alpha^c}Z_{b\alpha}^{ic}(B) = 0. \tag{28}$$

Further we shall consider that the basic density $\eta(x)$ of the space $M_n$ is connected with the gauge fields $B(x)$ by the following equations:

$$Z_{b\alpha}^{kc}(B)\frac{\partial \ln \eta}{\partial B_\alpha^c} + \xi_b^k = 0. \tag{29}$$

We shall take account by the making of the gauge field's equations that the total Lagrangian $\Lambda_t$ depends on covariant derivatives of these fields not higher of the first order. Let us to consider the variation of the total action (25) in regard on the following arbitrary infinitesimal substitutions:

$$x \to x, \quad \Psi \to \Psi, \quad \nabla_i \Psi \to \nabla_i \Psi, \quad B_\alpha^c \to B_\alpha^c + \delta_\circ B_\alpha^c, \quad \nabla_i B_\alpha^c \to \nabla_i B_\alpha^c + \delta_\circ \nabla_i B_\alpha^c,$$

where $\delta_\circ\left(\nabla_i B_\alpha^c\right) = \nabla_i\left(\delta_\circ B_\alpha^c\right) - T_i \delta_\circ B_\alpha^c$. By this the changes of fields functions $\delta_\circ B_\alpha^c(x)$ must equal to zero on the boundary $\Omega_{n-1}$ of the domain $\Omega_n$. As a result the action variation is written as:

$$\delta_\circ A_t = \int_{\Omega_n}\left(\frac{\partial(\Lambda_t \eta)}{\partial B_\alpha^c}\delta_\circ B_\alpha^c + \frac{\partial(\Lambda_t \eta)}{\partial\left(\nabla_i B_\alpha^c\right)}\delta_\circ\left(\nabla_i B_\alpha^c\right)\right)d_{\circ n}V =$$

$$\int_{\Omega_n}\left(\frac{1}{\eta}\frac{\partial(\Lambda_t \eta)}{\partial B_\alpha^c} - \nabla_i\left(\frac{\partial(\Lambda_t)}{\partial\left(\nabla_i B_\alpha^c\right)}\right)\right)\delta_\circ B_\alpha^c d_n V + \int_{\Omega_n}\left(\nabla_i\left(\frac{\partial \Lambda_t}{\partial\left(\nabla_i B_\alpha^c\right)}\delta_\circ B_\alpha^c\right) - T_i\frac{\partial \Lambda_t}{\partial\left(\nabla_i B_\alpha^c\right)}\delta_\circ B_\alpha^c\right)d_n V. \tag{30}$$

We transform the last integral in the formula (30) on the Green theorem (7) in the surface integral

$$\oint_{\Omega_{n-1}}\frac{\partial \Lambda_t}{\partial\left(\nabla_i B_\alpha^c\right)}\delta_\circ B_\alpha^c N_i \varepsilon(N)d_{n-1}V,$$

which becomes zero on account of the equality to zero of changes $\delta_\circ B_\alpha^c$ of gauge fields $B(x)$ on this surface. Thus for arbitrary changes $\delta_\circ B_\alpha^c$ the variation $\delta_\circ A_t$ of the total action $A_t$ can becomes zero only then if there are the following equations of gauge fields:

$$\Lambda_t\frac{\partial \ln \eta}{\partial B_\alpha^c} + \frac{\partial \Lambda_t}{\partial B_\alpha^c} - \nabla_i\left(\frac{\partial \Lambda_t}{\partial \nabla_i B_\alpha^c}\right) = 0. \tag{31}$$

Hence, contracting the expression (31) with $Z_{b\alpha}^{kc}(B)$, we receive

$$\nabla_i H_b^{ik} - S_{ij}^k H_b^{ij} = I_b^k,\tag{32}$$

where it are introduced the symbols:

$$I_b^k = -\Lambda_t \xi_b^k - \frac{\partial \Lambda_t}{\partial \nabla_k \Psi} X_b(\Psi) - \frac{\partial \Lambda_t}{\partial \nabla_k B_\alpha^c} Y_{b\alpha}^c(B),\tag{33}$$

$$H_a^{ij} = -H_a^{ji} = \frac{\partial \Lambda_t}{\partial \nabla_i B_\gamma^c} Z_{a\gamma}^{jc}(B), \qquad Y_{b\alpha}^c(B) = U_{b\alpha}^c(B) + T_i Z_{b\alpha}^{ic}(B).$$

We note that here it were used the correlations (28).
In adopted notation the correlations (26) having the form

$$\nabla_i I_b^i - T_i I_b^i = \frac{1}{2} H_c^{ik} R_{ikb}{}^c - H_b^{ik}\left(\nabla_i T_k + S_{ik}^j T_j\right) +$$

$$\left(\frac{\partial \Lambda_t}{\partial \Psi} - \nabla_i\left(\frac{\partial \Lambda_t}{\partial \nabla_i \Psi}\right)\right) X_b(\Psi) + \left(\frac{1}{\eta}\frac{\partial(\Lambda_t \eta)}{\partial B_\alpha^c} - \nabla_i\left(\frac{\partial \Lambda_t}{\partial \nabla_i B_\alpha^c}\right)\right) Y_{b\alpha}^c(B) \tag{34}$$

can be named the generalized differential laws of the conservation. Let us to note that the equations (34) are satisfied identically on extremals (10) and (31). The offered formalism can consider also and as the solutions (33) construction of differential equations (34).

## 5. Polarization fields

Further we concretize Greek indices replacing them by Latin indices *a, b, c, d, e, f, g, h* so that the components $B(x)$ of gauge fields will be noted in the form $B_a^c(x)$.

Probably the rank of the density matrix $\rho$ equals $n$, but it is impossible to eliminate that the generally given equality is satisfied only approximately when some components of a density matrix can be neglected. In any case we shall consider that among fields $B_a^b$ the mixtures $\Pi_a^i$ were formed with non-zero vacuum means $h_a^i$ which determine differentiable vector fields $\xi_a^i(x)$ for considered domain $\Omega_n$ as:

$$\Pi_a^i = B_a^b \xi_b^i \tag{35}$$

(fields $\xi_a^i(x)$ determine a differential of a projection $d\pi$ from $\Omega_r \subset M_r$ in $\Omega_n$). It allows to define a Riemannian space-time $M_n^\circ$, the basic tensor $g_{ij}(x)$ of which we shall introduce through a reduced density matrix $\rho'(x)$. As a result it can be "to hide" a part of fields by a non-trivial geometrical structure.

So let components $\rho_i^j$ of a reduced density matrix $\rho'(x)$ are determined by the way:

$$\rho_i^{\ j} = \xi^{+\,a}_{\ \ i}\,\rho_a^{\ b}\,\xi_b^{\ j}\Big/\!\left(\xi^{+\,c}_{\ \ k}\,\rho_c^{\ d}\,\xi_d^{\ k}\right) = \Pi^{+\,a}_{\ \ i}\,\Pi_a^{\ j}\Big/\!\left(\Pi^{+\,b}_{\ \ k}\,\Pi_b^{\ k}\right)\,, \qquad (36)$$

and let fields

$$g^{ij} = \eta^{k(i}\,\rho_k^{\ j)}\left(g^{lm}\,\eta_{lm}\right) \qquad (37)$$

are components of a tensor of a converse to the basic tensor of the space-time $M_n^\circ$. By this components $g_{ij}(x)$ of the basic tensor must be the solutions of following equations:

$$g^{ij}\,g_{ik} = \delta_k^{\ j} \qquad (38)$$

(hereinafter $\eta_{ij}$ are metric tensor components of a tangent space to $M_n^\circ$ and $\eta^{ik}$ are determined as the solution of equations: $\eta^{ij}\eta_{ik} = \delta_k^{\ j}$; $\delta_j^{\ i}$ are Kronecker deltas).

Let us note the integral (3) in the following manner

$$A_t = \int\limits_{\Omega_n} \Lambda_t d_n V = \int\limits_{\Omega_n}\left[\Lambda_\circ(B) + \Lambda_1(\Psi)\right]d_n V\,, \qquad (39)$$

where

$$\Lambda_1 = \kappa\,\overline{X^b}(\Psi)\rho_b^{\ a}\,X_a(\Psi) = \kappa\,\overline{D^a\Psi}\,D_a\Psi\Big/\!\left(B^{+\,c}_{\ \ b}\,B_c^{\ b}\right)\,, \qquad (40)$$

$$D_a\Psi = -B_a^{\ c}\,X_c(\Psi) = B_a^{\ c}\left(\xi_c^{\ i}\nabla_i\Psi - L_c\Psi\right)\,. \qquad (41)$$

Let the fields $D_a\Psi$ change analogously to the fields $\Psi(x)$ in a point $x \in M_n$, then is

$$\delta_\circ D_a\Psi = \delta\omega^b\left(L_b\,D_a\Psi - L_{ba}^{\ \ c}\,D_c\Psi - \xi_b^{\ i}\,\nabla_i D_a\Psi\right)\,, \qquad (42)$$

where the fields $L_{ba}^{\ \ c}(x)$ satisfy relations

$$L_{ac}^{\ \ d}L_{bd}^{\ \ e} - L_{bc}^{\ \ d}L_{ad}^{\ \ e} + \xi_a^{\ i}\nabla_i L_{bc}^{\ \ e} - \xi_b^{\ i}\nabla_i L_{ac}^{\ \ e} - \xi_a^{\ i}\xi_b^{\ j}R_{ijc}^{\ \ \ e} = -C_{ab}^{\ \ d}L_{cd}^{\ \ e}\,,$$

As a result we shall have

$$Y_{ca}^{\ \ d}(B) = C_{cb}^{\ \ d}B_a^{\ b} - L_{ca}^{\ \ b}B_b^{\ d} - \xi_c^{\ i}\nabla_i B_a^{\ d}\,, \qquad Z_{ba}^{kc}(B) = \delta_b^{\ c}\Pi_a^{\ k} \qquad (43)$$

($\delta_a^{\ b}$ are the Kronecker symbols).

Since the action (39) must be invariant by infinitesimal substitutions of the Lie local loop $G_r$, then the Lagrangian $\Lambda_\circ(B)$ must depend on the gauge [10] (boson) fields $B(x)$ by intensities $F_{ab}^c(B)$, having the form

$$F_{ab}^c = \Theta_d^c\left(\Pi_a^i\nabla_i B_b^d - \Pi_b^i\nabla_i B_a^d + \Xi_{ab}^d\right),\tag{44}$$

where

$$\Theta_b^c = \delta_b^c - \xi_b^i\Pi_i^a\left(B_a^c - \beta_a^c\right),\quad \Xi_{ab}^e = B_d^e\left(B_a^c L_{cb}^d - B_b^c L_{ca}^d\right) - B_a^c B_b^d C_{cd}^e\ .\tag{45}$$

Hereinafter a selection of fields $\Pi_i^a$ and $\beta_c^a$ are limited by the relations:

$$\Pi_j^a\Pi_a^i = \delta_j^i\ ,\qquad \beta_c^a\,\xi_a^i = h_c^i\ .\tag{46}$$

Further it is convenient to use the following Lagrangian:

$$\Lambda_\circ = \frac{\kappa_\circ'}{4}F_{ab}^c F_{ge}^d\left[t^{ag}\left(s_c^e s_d^b - \upsilon s_c^b s_d^e\right) + t^{be}\left(s_d^a s_c^g - \upsilon s_c^a s_d^g\right) + u_{cd}\left(t^{ag}t^{be} - \upsilon t^{ab}t^{ge}\right)\right]\tag{47}$$

($\kappa_\circ'$, $\upsilon$ are constants) [10]. If $s_a^b = \delta_a^b$, $t^{ab} = \eta^{ab}$, $u_{ab} = \eta_{ab}$ ($\eta_{ab}$ are metric tensor components of the flat space and $\eta^{ab}$ are tensor components of a converse to basic one) then the given Lagrangian is most suitable one at the description of the hot matter (all matter states are equally likely), because it is most symmetrical one concerning intensities of the gauge fields $F_{ab}^c$. What is more we shall require the realization of the correlations:

$$L_{cd}^{\ a}\eta^{db} + L_{cd}^{\ b}\eta^{da} = 0\,,\tag{48}$$

that the transition operators $L_{ac}^{\ b}$ generate the symmetry which follows from the made assumptions.

Let's connect non-zero vacuum means $\beta_a^b$ of gauge fields $B_a^b$ with a spontaneous violation of a symmetry which we must consider as a phase transition with a formation a Bose condensate from fermions pairs. The transition to the stage of the matter evolution of the observable region of the space for which one it is possible to suspect the presence of cluster states of weakly interacting particles will be expressed in following formula for tensors $s_a^b$, $t^{ab}$, $u_{ab}$ and $h_i^a$:

$$s_a^b = s\xi_a^i h_i^b + \xi_c^e\varepsilon_{\underline{e}}^b,\qquad t^{ab} = t\varepsilon_{(l)}^a\varepsilon_{(k)}^b\eta^{(l)(k)} + \varepsilon_{\underline{c}}^a\varepsilon_{\underline{d}}^b\eta^{\underline{cd}},$$

$$u_{ab} = u\xi_a^i\xi_b^j h_i^c h_j^d\eta_{cd} + \xi_a^c\xi_b^d\eta_{\underline{cd}},\qquad h_i^a = h_i^{(k)}\varepsilon_{(k)}^a\,.\tag{49}$$

$((i),(j),(k),(l),\ldots = 1,2,\ldots,n; \underline{a},\underline{b},\underline{c},\underline{d},\underline{e} = n+1, n+2,\ldots,n+\underline{r};\ \underline{r}\,/\,r \ll 1)$, where fields $h_i^{(j)}(x)$, taking into account the relations (49), are determined uniquely from equations: $h_k^a h_a^i = \delta_k^i$. Similarly tensors $\eta^{(i)(j)}$, $\eta^{\underline{ab}}$ are determined from equations: $\eta^{(i)(k)}\eta_{(j)(k)} = \delta_{(j)}^{(i)}$, $\eta^{\underline{ab}}\eta_{\underline{cb}} = \delta_{\underline{c}}^{\underline{a}}$, while tensors $\eta_{(i)(j)}$, $\eta_{\underline{ab}}$ are determined as follows: $\eta_{(i)(k)} = \eta_{ab}\,\varepsilon_{(i)}^{a}\,\varepsilon_{(k)}^{b}$, $\eta_{\underline{ab}} = \eta_{cd}\,\varepsilon_{\underline{a}}^{c}\,\varepsilon_{\underline{b}}^{d}$. We shall connect constants $\varepsilon_{(i)}^{a}$, $\varepsilon_{\underline{b}}^{a}$ with a selection of the gauge fields $\Pi_{i}^{a}(x)$ recording them by in the form

$$\Pi_{i}^{a} = \varepsilon_{(k)}^{a}\Phi_{i}^{(k)} + \varepsilon_{\underline{b}}^{a}\mathrm{P}_{i}^{b} \tag{50}$$

and let $\varepsilon_{\underline{b}}^{a} = 0$. Besides we shall apply the decomposition of fields $B_{b}^{a}(x)$ in the form

$$B_{c}^{a} = \zeta_{i}^{a}\Pi_{c}^{i} + \zeta_{\underline{b}}^{a}A_{c}^{b} \ , \tag{51}$$

where $A_{c}^{b} = \xi_{a}^{b}B_{c}^{a}$.

Note that we decompose the physical system described by fields $B_{b}^{a}(x)$ on two subsystems. One of them described by fields $\Pi_{a}^{i}(x)$, will play the role of the slow subsystem. In addition components of intermediate tensor fields $\xi_{a}^{i}(x)$, $\xi_{a}^{b}(x)$, $\zeta_{i}^{a}(x)$, $\zeta_{\underline{b}}^{a}(x)$ should be connected by the relations: $\zeta_{i}^{a}\,\xi_{a}^{j} = \delta_{i}^{j}$, $\zeta_{i}^{a}\,\xi_{\underline{a}}^{b} = 0$, $\zeta_{\underline{b}}^{a}\,\xi_{a}^{j} = 0$, $\zeta_{\underline{b}}^{a}\,\xi_{\underline{a}}^{c} = \delta_{\underline{b}}^{c}$. That's precisely what will be the first step by the construction of the condensed description for the present stage of the matter evolution of the Universe observable region. So, considering the indistinguishability of physical states of weakly interacting particles we shall use the reduced set of fields $\left\{\Pi_{c}^{i}(x), A_{c}^{b}(x)\right\}$ instead of the full set $\left\{B_{c}^{a}(x)\right\}$. Naturally, it is necessary to take into account that the constants performing the role of weighting coefficients such as $1/G_{N}$ ($G_{N}$ is the gravitational constant) appear in the Lagrangian.

## 6. The vector boson propagator

Let $n = 4$, $\upsilon = 2$, $tu = s^{2}$, $L_{c(k)}^{a} = L_{cb}^{\ a}\varepsilon_{(k)}^{b} = L_{c(k)}^{(i)}\varepsilon_{(i)}^{a}$, $L_{i(j)}^{(k)} = \zeta_{i}^{a}L_{a(j)}^{(k)}$, $L_{\underline{b}(j)}^{(k)} = \zeta_{\underline{b}}^{a}L_{a(j)}^{(k)}$, and so that the full Lagrangian (39) will be rewritten as follows

$$\begin{aligned}
\Lambda_{t} &= \Lambda(\Psi, D\Psi) + \frac{1}{4}\eta^{(j)(m)}\Big[\kappa_{\circ}\eta_{\underline{ab}}\eta^{(i)(k)}E_{(i)(j)}^{a}E_{(k)(m)}^{b} + \\
&\quad \kappa_{1}\Big(\eta_{(k)(n)}\eta^{(i)(l)}F_{(i)(j)}^{(k)}F_{(l)(m)}^{(n)} + 2F_{(i)(j)}^{(k)}F_{(k)(m)}^{(i)} - 4F_{(i)(j)}^{(i)}F_{(k)(m)}^{(k)}\Big)\Big],
\end{aligned} \tag{52}$$

where

$$\kappa_{\circ} = \kappa_{\circ}{}'t^{2}, \quad \kappa_{1} = \kappa_{\circ}{}'ts^{2}, \tag{53}$$

$$\begin{aligned}
F_{(i)(j)}^{(k)} &= F_{ab}^{c}\varepsilon_{(i)}^{a}\varepsilon_{(j)}^{b}h_{c}^{l}h_{l}^{(k)} = \Phi_{l}^{(k)}F_{mn}^{l}\Phi_{(i)}^{m}\Phi_{(j)}^{n} = A_{(i)}^{c}L_{c(j)}^{(k)} - A_{(j)}^{c}L_{c(i)}^{(k)} \\
&\quad + \Phi_{m}^{(k)}\Big(\Phi_{(i)}^{l}\nabla_{l}\Phi_{(j)}^{m} - \Phi_{(j)}^{l}\nabla_{l}\Phi_{(i)}^{m}\Big) + \Phi_{(i)}^{l}L_{l(j)}^{(k)} - \Phi_{(j)}^{l}L_{l(i)}^{(k)} \ ,
\end{aligned} \tag{54}$$

$$E_{ij}^a = E_{(k)(l)}^a \Phi_i^{(k)} \Phi_j^{(l)} = \xi_d^a F_{bc}^d \varepsilon_{(k)}^b \varepsilon_{(l)}^c \Phi_i^{(k)} \Phi_j^{(l)} =$$
$$\nabla_i A_j^a - \nabla_j A_i^a + C_{\underline{bc}}^a A_i^b A_j^c + C_{i\underline{b}}^a A_j^b - C_{j\underline{b}}^a A_i^b + C_{ij}^a \,, \tag{55}$$

$$\Phi_{(i)}^k = \Pi_a^k \varepsilon_{(i)}^a, \qquad A_i^b = A_c^b \Pi_i^c = A_{(j)}^b \Phi_i^{(j)}, \tag{56}$$

$$C_{\underline{ab}}^c = \xi_g^c C_{ed}^g \zeta_{\underline{a}}^e \zeta_{\underline{b}}^d \,, \; C_{i\underline{a}}^c = \xi_e^c \Big( C_{bd}^e \zeta_i^b \zeta_{\underline{a}}^d + \nabla_i \zeta_{\underline{a}}^e \Big) \,, \; C_{ij}^a = \xi_c^a \Big( C_{bd}^c \zeta_i^b \zeta_j^d + \nabla_i \zeta_j^c - \nabla_j \zeta_i^c \Big) \,. \tag{57}$$

As a result the equations of fields $\Phi_{(j)}^i(x)$ may be received in a standard manner [31] as the Einstein gravitational equations

$$R_{jik}{}^j - \frac{1}{2} g_{ik} g^{lm} R_{jlm}{}^j = \frac{1}{2\kappa_1} \Bigg[ g_{kj} \frac{\partial \Lambda}{\partial D_j \Psi} D_i \Psi - g_{ik} \Lambda$$
$$+ \kappa_\circ \eta_{\underline{ab}} g^{jl} \Big( E_{ij}^a E_{kl}^b - \frac{1}{4} g_{ik} g^{mn} E_{mj}^a E_{nl}^b \Big) \Bigg]. \tag{58}$$

( $R_{ijk}{}^l$ is the curvature tensor of the connection $\Gamma_{ij}^k$ of Riemannian space-time $M_n^\circ$; $\kappa_\circ = 1/(4\pi)$, $\kappa_1 = 1/(16\pi G_N)$ ). Naturally, that the Einstein equations only show a physical state of a matter. All this confirms a capability for interpretations of fields $\Phi_{(j)}^i(x)$ or fields $\Phi_i^{(j)}(x)$ as gravity potentials, but taking into account their dependence from properties of medium (vacuum), and also the opinion to consider components $g_{ij}(x)$ of a metric tensor of the space-time by potentials of a gravitational field, it is meaningful to call $\Phi_{(j)}^i(x)$ and $\Phi_i^{(j)}(x)$ by polarization fields. Precisely these fields describing the slow subsystem it can hide introducing the Riemannian structure of space-time, thereby we receive a possibility to apply the methods of differential geometry by the condensed description of physical systems not only in the cosmology, but and in the elementary particle physics [32].

Let's study an approaching, in which the space-time is possible to consider as a Minkowski space, the fields $\Phi_i^{(k)}$, $\Phi_{(k)}^i$ are constants and let $\underline{r} = 1$, that assumes $C_{\underline{ab}}^c = 0$. For obtaining equations of fields $A_i^b(x)$ in Feynman perturbation theory the calibration should be fixed. For this we shall add the following addend:

$$\Lambda_q = \frac{\kappa_\circ}{2} q_{\underline{bb}} g^{ij} g^{kl} \Big( \partial_i A_j^b - q_\circ C_i A_j^b \Big) \Big( \partial_k A_l^b - q_\circ C_k A_l^b \Big) \,, \tag{59}$$

to the Lagrangian (52), where $q_\circ = \eta_{\underline{bb}} / q_{\underline{bb}}$, $C_i = C_{i\underline{b}}^b$. Besides let

$$T_{a(k)}^{(i)} \eta^{(j)(k)} + T_{a(k)}^{(j)} \eta^{(i)(k)} = \varepsilon_a^b t_{\underline{b}} \eta^{(i)(j)} \,. \tag{60}$$

As a result of this equations of a vector field $A_i^b(x)$ will be written as:

$$g^{jk} \left[ \partial_j \partial_k A_i^a - (1 - 1/q_\circ) \partial_i \partial_j A_k^a + (1 - q_\circ) C_i C_j A_k^a \right] + m^2 A_i^a = I_i^a / \kappa_\circ, \qquad (61)$$

where $I_i^a = \dfrac{g_{ij}}{\eta_{\underline{aa}}} \dfrac{\partial \Lambda(\Psi)}{\partial A_j^a}$ and

$$m^2 = (n-1)(n-2) \kappa_1 t_{\underline{a}}^2 \left/ \left( 2 \kappa_\circ \eta_{\underline{aa}} \right) \right. - g^{jk} C_j C_k \ . \qquad (62)$$

Notice that owing to the vacuum polarization ($C_i \neq 0$) the propagator of a vector boson has the rather cumbersome view [32]

$$D_{ij}(p) = \left( p^m p_m - m^2 \right)^{-1} \left[ - g_{ij} + \right.$$
$$\left. (1 - q_\circ) \frac{\left( p_i p_j - C_i C_j \right) \left( p^k p_k - q_\circ m^2 \right) + (1 - q_\circ) p^k C_k \left( p_i C_j + C_i p_j \right)}{\left( p^l p_l - q_\circ m^2 \right)^2 + (1 - q_\circ)^2 \left( p^l C_l \right)^2} \right], \qquad (63)$$

which is simplified and receives the familiar form $\left( - g_{ij} / \left( p^k p_k - m^2 \right) \right.$, $p^k$ is the 4-momentum, and $m$ is the mass of the vector boson) only in the Feynman calibration ($q_\circ = 1$). This propagator allows to construct the renormalizable quantum theory of weak interactions ($D_{ij}(p) \to 0$, by $p \to \infty$), not attracting hypothetical scalar fields (the search of Higgs scalar bosons, forecasted in the standard model of electroweak interactions, is unsuccessful one for quite more quarter of a century). As a result it can make the conclusion that elementary particles must be considered right from the start as the compound physical systems for the correct description of which's it is necessary to attract polarization fields too. Polarization fields may be interpreted as fields describing "coat" consisting of virtual particles and surrounding the original bare particle.

## 7. Conclusion

We suppose that the greater part of weakly interacting particles constituting the great background of Universe exist in the degenerate (basic) state inserting the minor contribution in the vacuum polarization for the estimation of which the space curvature is used. In consequence of the low temperature $T_\circ$ of the Universe matter ground state (the density of the Universe matter ground state $n_\circ \sim 10^{-3} \, GeV^3$) the excited states of color fermions – the quarks in the form of baryons are distributed inhomogeneously and with a marginal density $\rho_b \sim 10^{-48} \, GeV^4$ ($\rho_b$ is the energy density of the Universe baryon matter). Therefore the geometrical structure of the space is distinguished from the structure of the flat space no too distinct. What is more it can assume that the symmetry of the Minkowski space-time is induced by physical properties of Universe fermions in a degenerate state when $T_\circ = 0$. It allows to solve not only the problem of the Universe planeness [1] but also to solve the problem of the observer horizon (the isotropy problem of the cosmic microwave background from the observer horizon of the Universe [1]).

The detection and the research of the neutrinos background of the Universe are the attractive problems. These problems do not seem the unpromising one in the case of the high neutrinos density. The main idea is it now for us what the normal matter (not neutrinos) acts as the Brownian particles by the help of which it can make the attempt to receive the estimation of parameters characterizing of the Universe dark matter. In the first place the use of the beta decay in all modifications suggests oneself [1]. Here there are the results raising us hopes by the work with the tritium, when by the search of the neutrino mass in most experiments it was detected the plateau instead of the energy spectrum dip, what was allows to advance a hypothesis on the presence of the appreciable neutrino halo within the limits of the baryon protocloud of Solar System [1]. Moreover, in like experiments it was received estimations of variations of the tritium decay period [33]. In consequence of this it gains in importance the further monitoring of parameters characterizing the weak interaction.